\documentclass[preprint,aps]{revtex4}
\usepackage{color}
\usepackage{mathtools}
\usepackage{diagbox}
\usepackage{tikz}
\usepackage{amsmath,amssymb,amsfonts,dcolumn,color,graphicx,graphics,latexsym,epsfig}
\usepackage[compatibility=false]{caption}
\usepackage{subcaption}
\usepackage{hyperref}
\usepackage{natbib}
\usepackage{float}
\usepackage{booktabs}
\usepackage{graphicx}
\usepackage{graphics}
\def\beq{\begin{equation}}
\def\eeq{\end{equation}}
\def\barr{\begin{array}}
\def\earr{\end{array}}

\newcommand{\tikzcircle}[2][red,fill=red]{\tikz[baseline=-0.5ex]\draw[#1,radius=#2] (0,0) circle ;}%
\begin{document}

\title{A new model with solitary waves: solution, stability and quasinormal modes}
\author{Surajit Basak ${}^{\dagger}$, Poulami Dutta Roy${}^\#$
and Sayan Kar ${}^{\#}$}
\email{surajit.basak@ifj.edu.pl, poulamiphysics@iitkgp.ac.in, sayan@phy.iitkgp.ac.in}
\affiliation{${}^{\dagger}$ Condensed Matter Physics, Institute of Nuclear Physics, Polish Academy of Sciences, ul. W. E. Radzikowskiego 152, PL-31342 Krak\'ow, Poland }
\affiliation{${}^\#$Department of Physics, Indian Institute of Technology 
Kharagpur, 721 302, India}

\begin{abstract}
\noindent We construct solitary wave solutions 
in a $1+1$ dimensional massless scalar ($\phi$) field theory with a
specially chosen potential $V(\phi)$. The equation governing perturbations 
about this solitary wave has an effective potential which is a simple harmonic
well over a region, and a constant beyond. 
This feature allows us to ensure the stability of the solitary wave through the existence of bound states in the well, which can be found 
by semi-analytical methods.  A further check on stability
is performed through our search for quasi-normal modes (QNM)
which are defined for purely outgoing boundary conditions. 
The time-domain profiles of the perturbations and the
parametric variation of the QNM values are presented and
discussed in some detail. Expectedly, a damped oscillatory
temporal behaviour (ringdown) of the fluctuations is clearly seen through our analysis of the
quasi-normal modes.

\end{abstract}


\maketitle

\newpage

\section {\bf Introduction}

\noindent The first observation of solitary waves dates back to 1834 when English naval architect John Scott Russell observed a wave in a water-canal which retained its shape while propagating over a large distance \cite{history}. Such unique solutions to the field equations of a specific theory were later found mathematically and 
are known as solitary waves. If a solitary wave retains its shape even after a collision with another of its kind, then they are 
called solitons.  Over the years,  solitary waves and solitons
have arisen in diverse branches of physics. Examples include optics \cite{rajaraman, optics}, cosmic strings \cite{cosmic string, cosmic string 1}, monopoles \cite{monopoles} superconductivity \cite{superconductor, superconductor1, superconductor2}, dark matter \cite{dark matter}, fluid mechanics \cite{fluid mechanics, fluid review}, antiferromagnetism \cite{antiferromagnetism}, particle physics \cite{particle physics, particle physics 1,Coleman} to name a few (see the cited review articles and references therein for further details on applications of solitons). Interestingly, a linear wave equation in any dimension can support non-dissipative as well as non-dispersive solutions which imply that the solutions retain their shape even if they travel from $x,t \rightarrow -\infty$ to $x,t \rightarrow \infty$. 
Such behaviour is evident from the dispersion relation where 
the angular frequency $\omega$ is a linear function of the 
wave number $k$. A simple scenario like this has a profound implication as it leads to the wave having equal phase and group velocity. Thus, all components of a wave packet corresponding to different frequencies travel with the same velocity keeping the shape of the wave packet intact, thereby producing a solitary wave. The situation changes drastically when an extra potential term
(representing an interaction) is added to the free wave equation. Depending on the nature of this extra term, the solution of the system may loose or retain its non-dissipative or non-dispersive property. 
Details on these
aspects which are directly linked to the ideas of solitary waves 
and solitons are nicely reviewed in \cite{Coleman,rajaraman}. 

\noindent For linear equations, if one assumes no dissipation, then to retain the shape of wave packets it is mandatory for the system to be non-dispersive as well. On the other hand, for non-linear systems with no dissipation, to keep the shape of the wave packets intact, the medium has to be dispersive. Hence a dispersive medium is essential
to balance the non-linearity and thus support a solitary wave.\\ 
Finding theories with non-linear equations of motion which support soliton/solitary wave solutions is easier said than done. There are some well known theories (e.g. $\phi^4- \phi^2$, sine-Gordon,
Korteweg-de Vries, Burgers ....
etc.) \cite{Coleman,kdv, rajaraman,sugiyama,lohe,bazeia,gomes,lensky,kink,gani2,oliveira} which do have such solutions. In general it can be said that any theory with a potential having multiple minima can support a solitary wave/soliton, interpolating between two neighbouring minima.
Solutions represented by a monotonically increasing function are sometimes called kinks
while those with a monotonically decreasing function are termed as anti-kinks. 
After finding such solutions the next task is to check if they have `particle-like' features. This is done by identifying a localised energy density, 
a finite total energy and a verification of its stability under
perturbations.

\noindent The interaction mechanisms between kink and antikink solutions vary with the solitary wave model. For instance, completely integrable solitary wave models (sine-Gordon model) have a simple interaction picture of kink-antikink collisions wherein they pass with just a phase shift or time delay.
The picture is a lot more interesting in models like $\phi^4$ theory (non-integrable models) where we get non-trivial interactions. The end state of collision can lead to annihilation (trapped/bound state) for low initial velocities or reflection corresponding to high velocities. Moreover, there are alternate ``windows'' of initial velocity called resonance windows that correspond to annihilation or reflection and have been well studied in the literature \cite{gani,gani1,decker,simas,campbell,resonance}. This interesting behaviour is called `resonance structure' which has been observed to be suppressed by the quasi-normal modes of the system as well \cite{campos,Dorey}.

\noindent In the following sections we will discuss a scalar
field theory with a novel potential, constructed in a way to support a solitary wave solution. Our work is organised as follows. In Sec.II  we construct our scalar field model with a tailor-made potential 
and show that it does support solitary waves. The potential has the desired double minima structure, the energy density is found to be localised and the total energy is finite and positive. 
The stability of the solitary waves is discussed in Sec.III. A confined harmonic oscillator potential 
arises  when we perturb the solitary wave. Thus there are normal modes
or bound states. We also explore the possible existence of transmission resonances 
and quasi-normal modes in the system for purely
`outgoing' boundary conditions. We end with a brief conclusion in Sec. IV. 
 
\section{\bf A new scalar field model with solitary waves}

\noindent Consider a scalar field theory in (1+1)- dimensions with a Lagrangian density given as
\begin{equation}
\mathcal{L}=\partial_{\mu}\phi\partial^{\mu}\phi-\tilde{V}(\phi)
\end{equation}
where $\phi$ is the scalar field having dependence on $t$ and $x$ and the 
potential $\tilde{V}$ is a function of $\phi$. The corresponding equation of motion, with $c=1$, is
\begin{equation}\label{eqm}
\frac{\partial^2 \phi(x,t)}{\partial t^2}-\frac{\partial^2 \phi(x,t)}{\partial x^2}=-\frac{d \tilde{V}}{d \phi}
\end{equation}
with the total energy  
\begin{equation}\label{energy}
E=\int_{-\infty}^{\infty} dx \Big[\frac{1}{2}\,\Big(\frac{\partial \phi}{\partial t}\Big)^2+\frac{1}{2}\,\Big(\frac{\partial \phi}{\partial x}\Big)^2+\tilde{V}(\phi)\Big] .
\end{equation}
For the total energy to be finite, or the energy density to be localised (by localised we mean it goes to zero as $x \rightarrow \pm \infty$), the integrand in (3) must go to zero as $x\rightarrow \pm \infty$. Ensuring the localization of energy 
density would be the first step in verifying the solitary wave nature of the solution. To achieve this, the derivatives of $\phi$ and the potential $\tilde{V}(\phi)$ must reach one of its zeroes as $x\rightarrow \pm \infty$. If the potential has a minimum corresponding to $\phi=0$ only, we have a trivial solution. Hence, for non-trivial solutions with localized energy density, the potential must have atleast two minima. In the next subsection we will specifically construct such a potential satisfying the required criteria. 

\noindent We mention that in this article,
we have chosen to work in natural units ($h = c =1$), 
which makes $\phi$ dimensionless. The potential $\tilde{V}(\phi)$ is still in units of inverse length squared. 
We can 
change our dimensionful variables $(x,t)$ to the 
dimensionless pair $x'= \sqrt{V_0}x$ and $t' = \sqrt{V_0}t$ where $V_0$ has dimensions of inverse length squared and 
may be associated with an overall scale for the
potential. This will be clearer when we formulate the potential explicitly in Sec. II(A) (see eq.(\ref{potential})). In such dimensionless space and time variables 
we need not worry about units.

\noindent Rewriting the total energy in terms of ($x',t'$) 
we have
\begin{equation}\label{energy_dimensionless}
E=\int_{-\infty}^{\infty} dx' \Big[\frac{1}{2}\,\Big(\frac{\partial \phi}{\partial t'}\Big)^2+\frac{1}{2}\,\Big(\frac{\partial \phi}{\partial x'}\Big)^2+V(\phi)\Big] .
\end{equation}
where $\tilde{V}(\phi) = V_0 V(\phi)$ making $V(\phi)$ a dimensionless quantity.
Proceeding with our aim of obtaining 
static solutions first, we note that the field equation
reduces to 
\begin{equation}\label{static_eq}
\frac{d^2 \phi(x')}{d x'^2}=\frac{d V}{d \phi}.
\end{equation}
Since the theory is relativistically invariant, once we find a localised solution of eq.(\ref{static_eq}), we can boost it to a frame where the same solution will appear to be 
propagating with an unchanged profile, thereby 
representing a solitary wave. Multiplying the earlier equation on both sides by $\frac{d\phi}{dx'}$, integrating, and remembering that $\frac{d \phi}{d x'}$ goes to one of the minima of $V(\phi)$ at $x' \rightarrow \pm \infty$, one arrives at a first order differential equation,
\begin{equation}\label{eq_for_V}
\frac{d\phi}{dx'}=\pm \sqrt{2 V(\phi)}.
\end{equation}
The solution to the equation (\ref{eq_for_V}) can be identified as $\phi_{k}(x')$. The R.H.S can either be always positive or always negative. Thus, for any solitary wave, the static solution is always monotonically increasing between two successive minima of the potential. The monotonically decreasing solution is called an anti-soliton.

\subsection{Constructing the potential $V(\phi)$}

\noindent In order to obtain solitary wave solutions in a scalar field theory, we need a suitable form of the potential V($\phi$). The novelty in our work is the use of a 
potential which is different from the known ones yielding
solitary waves. We follow a specific route in obtaining the
potential--the reason behind this way of arriving at $V(\phi)$ will become
clear later. Differentiating the static equation (\ref{static_eq}) with respect to $x'$ we get,
 \begin{equation}
 \frac{d^3 \phi_{k}}{dx'^3}=\frac{d^2 V(\phi_{k})}{d\phi^{2}_{k}}\frac{d\phi_{k}}{dx'}=U(x')\frac{d\phi_{k}}{dx'}
 \end{equation}
where all the $\phi's$ have been replaced by the static solution $\phi_{k}$ since we wish to have a $V(\phi)$ which will solve eq.(\ref{eq_for_V}) with $\phi_k(x')$. We demand $U(x')$ to be in the form of harmonic potential--this being the central idea behind our construction of $V(\phi)$. Replacing $U(x')$ by $\alpha(\alpha x'^2-1)$ and assuming $\frac{d\phi_{k}}{dx'}=z$ we obtain,
\begin{equation}
\frac{d^2 z}{dx'^2}= \alpha(\alpha x'^2-1)z.
\label{eq:z}
\end{equation}

\noindent The reason for such a specific demand and the role of $U(x')$ will be  evident in section III (see eq.(\ref{Schrodinger})) where we discuss
fluctuations. In short, the solvability of the fluctuation equation
is behind this choice. Similar attempts have been discussed
earlier in \cite{campos}, \cite{Dorey}.\\
\noindent A solution to the above equation (\ref{eq:z}) is $z= e^{-\frac{\alpha x'^2}{2}}$. By integrating further, we arrive at an expression for $\phi_{k}$ as
\begin{equation}
\phi_{k}\,=\,\int e^{-\frac{\alpha x'^2}{2}} dx'\,=\,\sqrt{\frac{\pi}{2 \alpha}}\, erf\Big(\sqrt{\frac{\alpha}{2}} x'\Big)
\end{equation}
where $erf(\sqrt{\frac{\alpha}{2}} x')$ is the error function. The above equation can be inverted to get an expression for $x'$ in terms of $\phi_{k}$, which, when compared with eq.(\ref{eq_for_V}) gives
\begin{equation}
\int \frac{d\phi_{k}}{\sqrt{2 V(\phi_{k})}} = x' =\sqrt{\frac{2}{\alpha}}\, erf^{-1} \Big(\sqrt{\frac{2 \alpha}{\pi}}\phi_{k}\Big)
\end{equation}
with $erf^{-1} (\sqrt{\frac{2 \alpha}{\pi}} \phi_{k})$ being the inverse error function. Differentiating the above equation with respect to $\phi_{k}$ we can get the expression for $V(\phi_{k})$
\begin{equation}
V(\phi_{k})=\frac{1}{2} Exp\Big[-2\Big[erf^{-1}\Big(\sqrt{\frac{2 \alpha}{\pi}}\phi_{k}\Big)\Big]^2\Big] .
\end{equation}

\noindent We can now remove the subscript `$k$' from $\phi$ and
have a general definition of the potential $V(\phi)$. However, such a function does not have any minimum and is not defined on the entire
real line. Hence it is not useful for constructing solitons or solitary waves. As a remedy, we have joined two quadratic functions at $\phi = \pm b$ which have well-defined minima at $\phi=a$ and $\phi=-a$ ($a>b$). We use the expression for $V(\phi_{k})$ to motivate the final form of the potential $V(\phi)$, which is given below. A plot of the potential is shown in Fig.(\ref{fig:pot}).
\begin{equation}\label{potential}
V(\phi)  = \begin{cases} \frac{1}{2}(\phi + a)^2 & ; \,\, -\infty <\phi \leq -b \\
\frac{1}{2} Exp[-2[erf^{-1}(p\phi)]^2] & ; \,\,-b \leq \phi \leq b \\
 \frac{1}{2}(\phi - a)^2 & ;  \,\,   b \leq \phi < \infty
\end{cases}
\end{equation}
with $ [erf^{-1}(p\phi)]^2 $ being the square of the inverse error function and $p$ an independent parameter introduced while  generalizing from $V(\phi_k)$ to $V(\phi)$ .
If one wishes to revert back to dimensional variables, $V(\phi)$ will be scaled up by factor of $V_0$, which has units of inverse length squared.

\begin{figure}[h]
     \centering
      \includegraphics[scale=0.98]{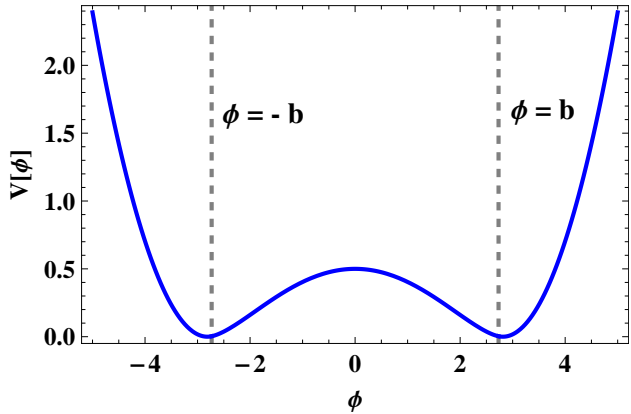}
      \caption{The potential having two minima at $\phi=\pm a$. The values of the parameters used in the plot are mentioned in eq.(\ref{parameters}).}
      \label{fig:pot}
\end{figure}
\noindent The potential is constructed in such a way that $V(\phi)=V(-\phi)$. While choosing the parameters, the continuity of both V and its derivative at $\phi=\pm b$ must be ensured. Thus, we are lead to the following conditions:
\begin{equation}
\left (b-a \right )^2 = Exp[-2[erf^{-1}(pb)]^2]
\label{eq:continuity}
\end{equation}
\begin{equation}
b-a=-\sqrt{\pi}\,p\, Exp[-[erf^{-1}(pb)]^2]\,erf^{-1}(pb).
\label{eq:derivative}
\end{equation}
Eq.(\ref{eq:continuity}) corresponds to the continuity of $V(\phi)$ at $\phi = \pm b$. While taking the square root in  eq.(\ref{eq:continuity}) we must remember that $a>b$. For eq.(\ref{eq:derivative}), the odd nature
of the inverse error function leads to the same equation for $\phi =\pm b$. Among the three independent parameters namely $a$, $b$, $p$, two can be fixed by the above equations while the other needs to be fixed by hand. An additional scaling parameter $V_0$ (having inverse length squared units) will appear in the expression for the potential, if one is working with dimensionful variables. It may seem that the introduction of
the parameter $p$ is superfluous. However, without $p$  (i.e. $p=1$), eqns. (\ref{eq:continuity}) and (\ref{eq:derivative}) completely fix $a$ and $b$ and the generality of the model is lost, which is not desirable.  Another fact to be noted is that our potential V($\phi$) has different functional dependencies on the field $\phi$ in different regions. Hence it will transform differently under rescaling of $\phi$
in different regions. Such a feature is not really forbidden
and can only be tested if, in  any way, our model becomes
useful in a real and experimentally realisable context. In addition,
this feature of $V(\phi)$ does explain the significance of the 
parameter $p$ introduced above. In the forthcoming discussions, we will use the following values for the parameters wherever necessary, if not specified otherwise.
\begin{equation}\label{parameters}
a=2.81354,\; b=2.73146,\; p=0.356825.
\end{equation}

\noindent There is
no specific advantage with the choice of these values and other
sets of values for the parameters can always be chosen. With the chosen values of all parameters, the equality of eq.(\ref{eq:continuity}) and eq.(\ref{eq:derivative}) is respected upto 6 decimal places.

\subsection{The solitary wave solution}
\begin{figure}[h]
 \centering
 \begin{subfigure}[t]{0.43\textwidth}
 	\centering
 	\includegraphics[width=\textwidth]{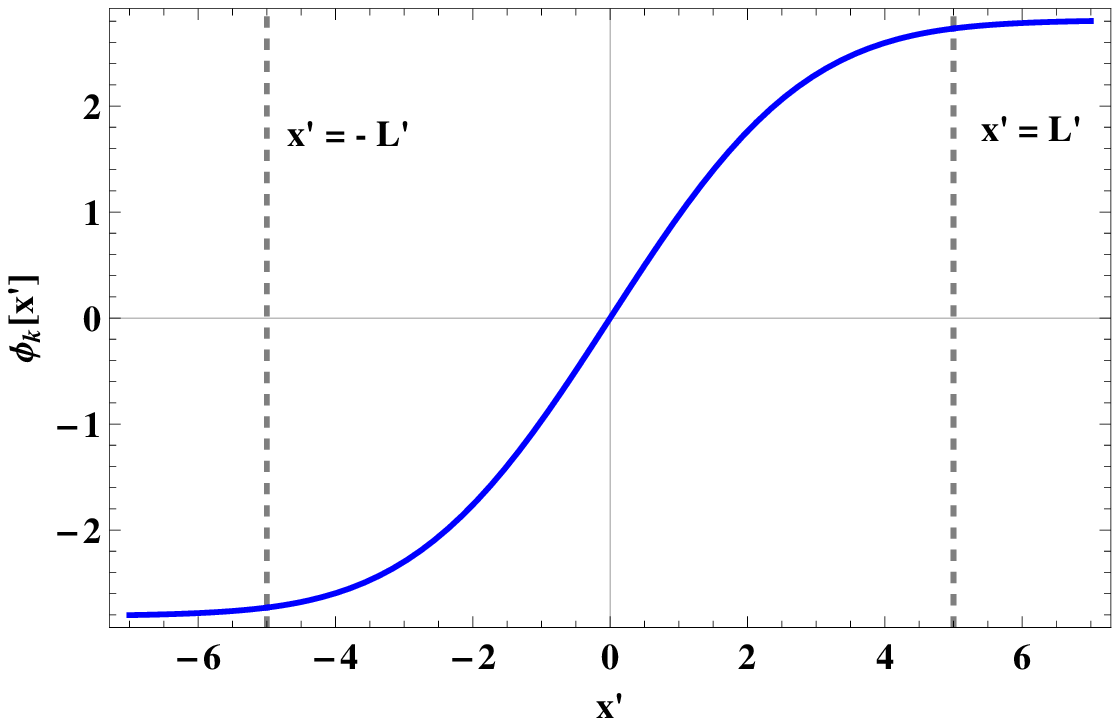}
 	\caption{Static solution starts from one minima of the potential $(\phi=-a)$ at $x'\rightarrow -\infty$ and ends at the other $(\phi=a)$ at $x' \rightarrow \infty$.}
 	\label{fig:solp}
 	\end{subfigure}
 	\hspace{0.2in}
 \begin{subfigure}[t]{0.43\textwidth}
 	\centering
 	\includegraphics[width=\textwidth]{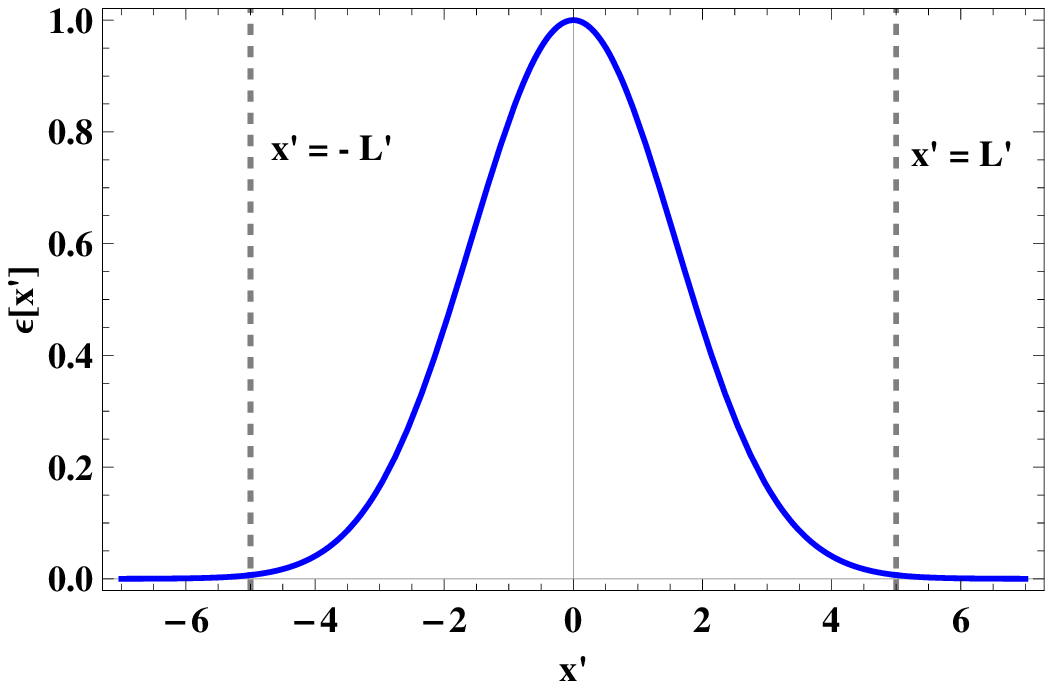}
 	\centering
 	\caption{ Energy density centred at $x'=0$ and dying off on both sides.}
 	\label{fig:eneg}
 	\end{subfigure}	
 	\caption{In figure (a) and (b) the dashed line denotes the point $x'=L'$. At $x'=L'$ in fig.(a) the value of $\phi$ is $b$. In all the figures, previously assumed values of the parameters (eq.(\ref{parameters})) has been used.}
 \end{figure}

\noindent With the specific form of the potential in hand, we can now proceed to solving the static equation for $\phi$ (eq.(\ref{eq_for_V})) with the form of the potential $V(\phi)$ given in eq.(\ref{potential}). The stationary solution for the specific ranges of $x'$ turns out to be of the following form
\begin{equation}
\phi  =\begin{cases} -a-(b-a)\,Exp\Big[\,(x'+L')\Big] &; \,\, -\infty < x' \leq -L' \\
\frac{1}{p}\,erf\Big(\frac{\sqrt{\pi}}{2}\,p\,x' \Big) &; \, \; -L' \leq x' \leq L' \\
a+(b-a)\,Exp\Big[-\,(x'-L')\Big] & ; \,\, L' \leq x' < \infty
\label{eq:phi_k}
\end{cases}
\end{equation}
where $x' = L'$ corresponds to the point $\phi = b$, when the potential $V$ is expressed in terms of $x'$. $x'=L'$ corresponds to $x=L$ with $L' = \sqrt{V_0}L$.  While constructing the stationary solution we need to keep in mind that our solitary wave must reach $\phi \rightarrow \pm a$ i.e. the points of potential minima, asymptotically. Therefore, the $(\pm)$ sign from the square root in eq.(\ref{eq_for_V}) has to be chosen such that $d \phi/ dx'$ is positive everywhere. In the range $L' \leq x' < \infty$ we take the negative sign because the square root of the potential turns out to be $ (\phi -a)$. But the values of $\phi$ begin from $\phi =b < a$. Thus, to keep the R.H.S.  positive we need to take the negative sign of the square root. Now one might wonder about the case when $ a \leq \phi < \infty$, where the quantity $(\phi -a)$ becomes positive again. This does not pose a problem because, though $\phi$ can go to $\infty$ as the variable in the potential, for the solitary wave, we require $\phi = \pm a$  at the infinities of $x'$. Therefore, $(\phi -a)$ will always be less than or equal to $0$ and hence the negative sign will keep $d\phi/ dx'$ positive. A similar argument will hold for the range $-\infty < x' \leq -L'$ where the positive sign of the square root in eq.(\ref{eq_for_V}) is suitable. \\
\noindent The value of $L'$ can be obtained by using the fact that $\phi = b$ when $x' = L'$. The form of $\phi_k$ in the range $-L'\leq x' \leq L'$ resembles the expression we had obtained earlier, with some additional factors arising due to a generalization from $V(\phi_k)$ to $V(\phi)$. The value of $L'$ for our choice of parameter values turns out to be,
\begin{equation}
b\,=\,\frac{1}{p}\,erf\Big(\frac{\sqrt{\pi}}{2}\,p\,L'\Big) \quad \Rightarrow L'=5.
\end{equation}
The form of the static solution as shown in Fig.(\ref{fig:solp}) confirms that it is indeed a solitary wave, interpolating between two neighbouring minima ($\phi=\pm a$) and monotonically increasing in between. The form of the anti-soliton can similarly be found from eq.(\ref{eq_for_V}) using the negative sign.

\noindent The energy density for the solitary waves can be obtained using  eq.(\ref{energy_dimensionless}) and eq.(\ref{eq_for_V}) which gives us 
the function, $2V(\phi_{k})$. The expression for energy density $\epsilon (x') $ as obtained for different ranges of $x'$, by substituting $\phi_k$ from eq.(\ref{eq:phi_k}) in $V(\phi_k)$, is centred at $x'=0$ and decays 
on either side. It is given as,  
\begin{equation}
\epsilon (x')  = \begin{cases} (b-a)^2\,Exp\Big[2(x'+L')\Big] &; \,\, -\infty < x' \leq -L' \\
 Exp\Big[-\frac{\pi}{2}p^{2}x'^2\Big] & ; \,\, -L' \leq x' \leq L' \\
(b-a)^2\,Exp\Big[-2 (x'-L')\Big] &; \,\, L'\leq x'<\infty.
\end{cases}
\end{equation}
Fig.(\ref{fig:eneg}) demonstrates the localized nature of the energy density of the solitary wave. Integrating, we find the total energy 
to be, 
\begin{gather}
\begin{split}
E & =\int_{-\infty}^{\infty} \, \, 2 V(\phi_k)\, dx'\\
& = 2 (b-a)^2 \int_{L'}^{\infty} Exp[-2  (x'-L')] dx' + \int_{-L'}^{L'} Exp \Big[\frac{-\pi p^2 x'^2 }{2} \Big]\, dx' \\
& =\frac{\sqrt{2}}{p}erf\Big(\sqrt{\frac{\pi }{2}}\,L'\,p\Big)+\,(b-a)^2 = 3.96386.
\end{split}
\end{gather}
Note that the value quoted above (for the chosen values of the parameters as mentioned in eq.(\ref{parameters})) is indeed positive and finite. The energy
density and total energy will be scaled up with $V_0$ and $\sqrt{V_0}$, respectively if we work with $(x,t)$ variables. In that case, energy will be in inverse length units.

\noindent The above discussion shows that with our chosen potential
we do end up with solitary waves having localised energy density and
finite total energy. In addition, by construction, we are assured
of the stability of the solution, as we now explain in detail
in the following section.

\section{\bf Stability of solitary waves}

\noindent As is well known, solitary waves exhibit 
particle-like behaviour, a feature manifest in the 
nature of the solution. We would now like to
investigate the stability of the solitary waves 
obtained in the previous section, under small time-dependent perturbations. In other words, we add a small perturbation $\eta(x') e^{i \omega' t'}$ to the static solution, i.e. 
\begin{equation}\label{eq:phi_perturb}
\phi(x',t')=\phi_{k}(x')+e^{i \omega' t'} \eta(x')
\end{equation}
where $\omega'$ is a dimensionless frequency related to the dimensionful frequency via $\omega'^2 = \frac{\omega^2}{V_0}$ with $V_0$ in units of inverse length squared. The total field $\phi(x',t')$ still has to obey the time dependent equation (\ref{eqm}), scaled appropriately. Substituting (\ref{eq:phi_perturb}) in eq.(\ref{eqm}), expanding $V(\phi)$ around the static solution $\phi=\phi_{k}$ and keeping terms only upto first order in $\eta$, we find,
\begin{equation}\label{Schrodinger}
\Big[-\frac{d^2}{dx'^2}+U(x')\Big]\,\eta(x')=\omega'^2\,\eta(x')
\end{equation} 
where $U(x')=\frac{d^2 V}{d\phi^2}|_{\phi=\phi_{k}}$. For the solitary wave to be stable under perturbation, the Schr\"{o}dinger-like equation (\ref{Schrodinger}) must admit solutions with positive and real values of $\omega'^2$ (normal modes).
\subsection{Confined harmonic oscillator as effective potential}

\begin{figure}[h]
      \centering
      \includegraphics[scale=1.1]{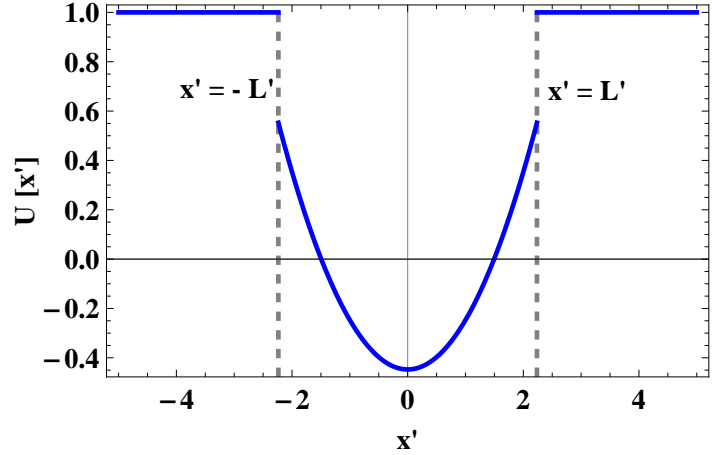}
      \centering
      \caption{ The harmonic oscillator potential confined between two walls of finite height. Here $\alpha=0.2$ and $L'=5$ following the values mentioned in eq.(\ref{parameters}).}
      \label{fig:upot}
  \end{figure}

\noindent From the way in which we constructed $V(\phi)$ in Sec.II (A), we already have $U(x')$ in the following form 
\begin{equation}\label{U(x)}
\frac{d^{2}V}{d\phi^2}\Big|_{\phi=\phi_{k}} = U(x') = \begin{cases} 1 &; \,\, -\infty< x' <-L' \\
 \alpha(\alpha x'^2-1) &; \,\, -L' \leq x' \leq L' \\
1   &; \,\; L' < x' < \infty
 \end{cases}
\end{equation}
where $\alpha = \frac{\pi}{2} \, p^2$ and $L'$ can be written as a function of $\alpha$ as $L' =1/\alpha$. The parameter
values chosen satisfy the conditions required for a solitary wave.
The plot in Fig.(\ref{fig:upot}) shows the harmonic
oscillator potential bounded by constant potential walls on both sides. It is interesting to note that even though $V(\phi)$ is continuous at the points $x' = \pm L'$, $U(x')$ is not. This is because of the fact that the quantity $\alpha$ is pre-determined from our requirement of a continuous $V(\phi)$. It can be checked that there are no values of parameters $p$, $a$, $b$ that can simultaneously make both $V(\phi)$ and $U(x')$ continuous. The nature of the potential $U(x')$ is
reminiscent of a square well, except that the depth of the well
now varies within the region of confinement.   

\noindent The general solution for $\eta(x')$, that dies off as $x' \rightarrow \pm \infty $ is 
\begin{equation}
\eta(x') =\begin{cases} A e^{-mx'} &; \,\,  L' \leq x'<\infty \\
 Bf(x')+Cg(x')&; \,\,  -L' \leq x' \leq L' \\
D e^{mx'} &; \,\, -\infty<x' \leq -L'
\end{cases}
\end{equation}
where $m=\sqrt{1-\omega'^2}$ and $f(x'),g(x')$ are the solutions of the equation
\begin{equation}
\Big[-\frac{d^{2}}{dx'^{2}}+\alpha\,(\alpha x'^2-1)\Big]\,\eta(x')=\omega'^{2}\,\eta(x')
\end{equation}
and are found to be, 
\begin{equation}\label{hypergeometric}
\begin{split}
f(x')=e^{\frac{\alpha x'^2}{2}} \,_1F_{1}\Big(\frac{1}{2}+\frac{\omega'^2}{4\alpha};\frac{1}{2};-\alpha x'^2\Big)\\
g(x')= x'e^{\frac{\alpha x'^2}{2}} \,_1F_{1}\Big(1+\frac{\omega'^2}{4\alpha};\frac{3}{2};-\alpha x'^2\Big)
\end{split}
\end{equation}
 with $\,_1F_1$ being the confluent hypergeometric function. Note that for the first solution $f(x')$ we have $f(x')=f(-x')$ and $f'(x')=-f'(-x')$ while the reverse is true for the second solution, i.e. $g(x')=-g(-x')$ and $g'(x')=g'(-x')$.\\
The continuity of $\eta(x')$ and its derivative at $x'=\pm L'$ along with the use of the above data, results in the following equations, 
\begin{equation}\label{continuity}
\begin{split}
 Bf(L')+Cg(L')= A e^{-mL'} \, \hspace{0.4in}
 Bf'(L')+Cg'(L')= -mA e^{-mL'} \\
Bf(L')-Cg(L')= D e^{-mL'} \, \hspace{0.4in}
-Bf'(L')+Cg'(L')= mD e^{-mL'} .
 \end{split}
\end{equation}
We have to find values of $\omega'^2$ for which the equations in (\ref{continuity}) are satisfied. After some straightforward algebra,
we arrive at
\begin{equation}\label{continuity1}
\begin{split}
2Bf(L')=(A+D)e^{-mL'} \qquad 2Bf'(L')=-m(A+D)e^{-mL'}\\
2Cg(L')=(A-D)e^{-mL'} \qquad 2Cg'(L')=-m(A-D)e^{-mL'}.
\end{split}
\end{equation}
From the first set of equations in (\ref{continuity1}) it is evident that if $B \neq 0$ and $A \neq -D$, we have
\begin{equation}\label{f(L)}
\frac{f'(L')}{f(L')}=-m.
\end{equation} 
Similarly, from the second set of equations we get a condition for $g(L')$ if $C \neq 0$ and $A \neq D$,
\begin{equation}\label{g(L)}
\frac{g'(L')}{g(L')}=-m.
\end{equation}
Both the equations (\ref{f(L)}) and (\ref{g(L)}) cannot be solved simultaneously for any real value of $m$. Therefore, we have to deal separately with the two cases at hand. Since the above equations for $\omega'^2$ are transcendental, they can only be solved graphically or numerically. For each equation, we first plot the LHS and RHS in the same grid using {\em Mathematica} and find the intersection points which are solutions to the corresponding equation. The exact numerical value can then be found using the {\em FindRoot} command in {\em Mathematica 10}. \\

\begin{figure}[h]
 \centering
\begin{subfigure}[t]{0.44\textwidth}
  \centering
	\includegraphics[width=\textwidth]{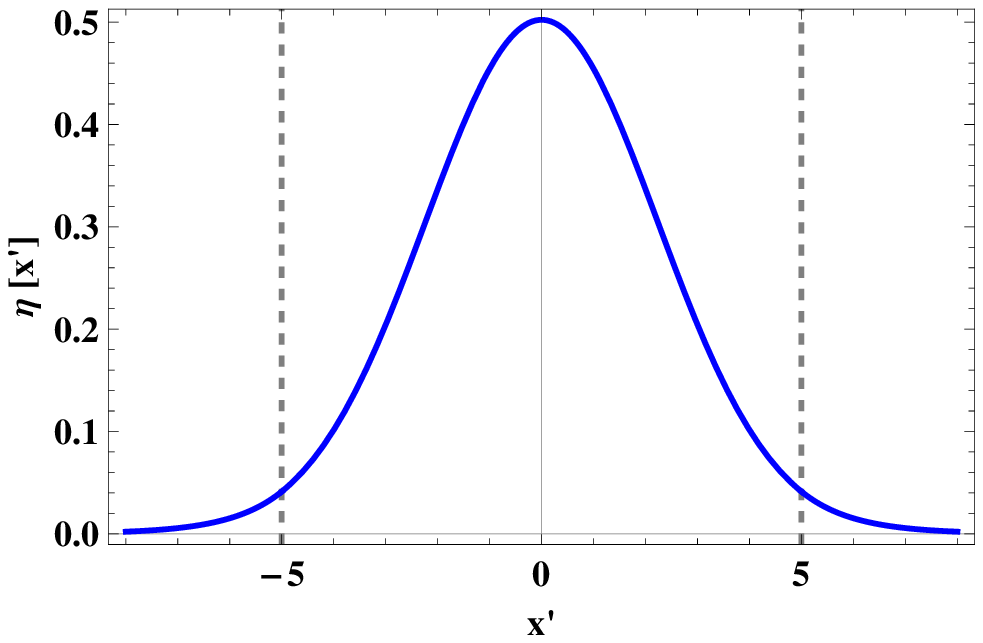}
 	\caption{The ground state($\omega'^2=0$)}
 \end{subfigure}
 \hspace{0.1in}
 \begin{subfigure}[t]{0.44\textwidth}
 	\centering
 	\includegraphics[width=\textwidth]{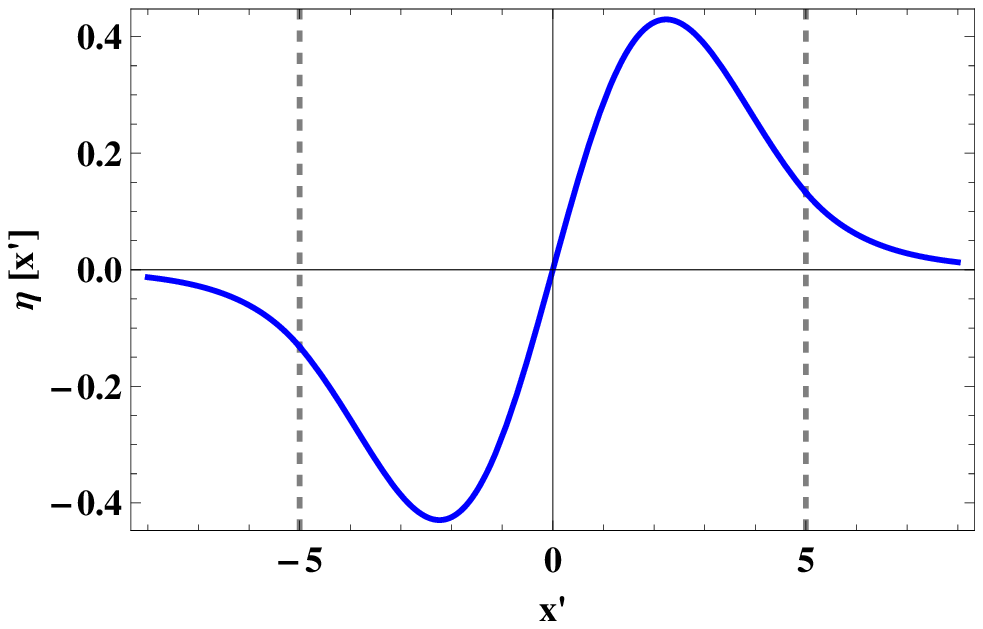}
 	\caption{ First excited state($\omega'^2=0.399125$)}
 	\end{subfigure}
 \begin{subfigure}[t]{0.44\textwidth}
 	\centering
 	\includegraphics[width=1.1\textwidth]{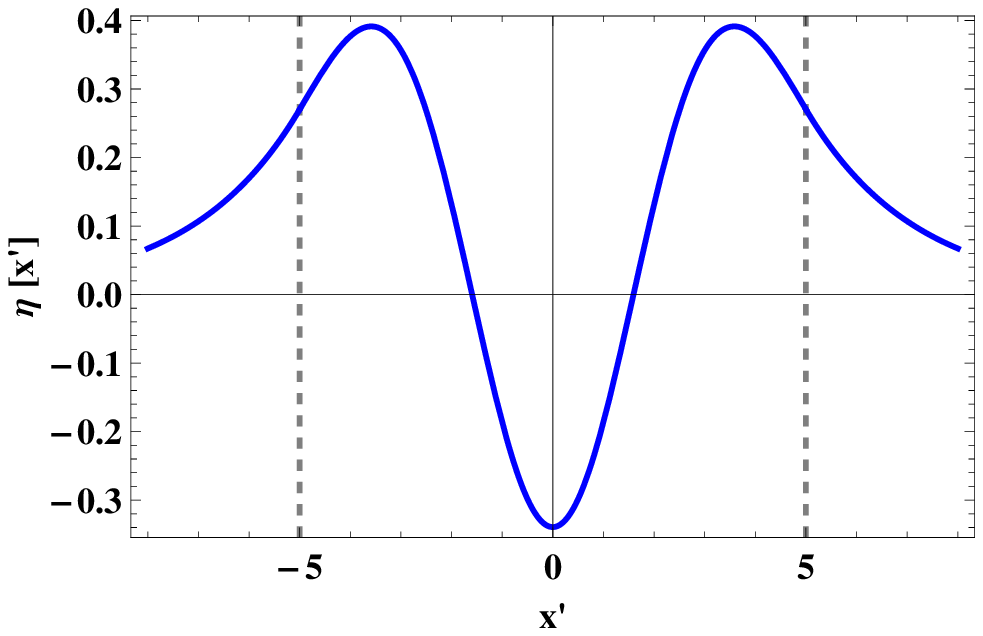}
 	\centering
 	\caption{ Second excited state($\omega'^2=0.785064$)}
 	\end{subfigure}	
 	\caption{Plots of wavefunctions for different energy eigenstates corresponding to $\alpha=0.2$, $L'=5$. Dashed lines mark the position of $L'$. Note that the increasing $\omega'^2$ solutions have more nodes, as expected from any solution of the Schr\"{o}dinger--like equation.}
 	\label{fig:wf}
 \end{figure}
\noindent \textbf{Case I: $B \neq 0$ , $A \neq -D$ $\&$ $C=0$ , $A=D$}\\
There are two solutions for $\omega'^2$ satisfying eq. (\ref{f(L)}), one with $\omega'^2=0$ $(m = 1)$ and the other 
with $\omega'^2=0.785064$ $(m = 0.463612)$. We have used the values of the parameters as given in eq.(\ref{parameters}). Adopting normalization
(not really a necessity for our purposes here), the constant coefficients can also be calculated from eq.(\ref{continuity}), and are found out to be $A=D=6.11895,\,B=0.50227$ and $A=D=2.74436,B= -0.33936$ for $\omega'^2=0$ and $\omega'^2=0.78506$ respectively.\\

\noindent \textbf{Case II: $C \neq 0$ , $A \neq D$ $\&$ $B=0$ , $A=-D$}\\
In this case a single solution can be found with $\omega'^2=0.399125$ $(m = 0.77516)$ and the coefficients are $A=-D=6.37876,\,C=0.316409$. The plots in Fig.(\ref{fig:wf}) shows the wave-functions for different 
$\omega'^2$.\\
From the above discussion 
it is apparent that there are no {\em negative $\omega'^2$} solutions to eq.(\ref{Schrodinger}). The normal modes of the system have frequency $\omega'$.

\subsection{Are there quasinormal modes?}
\noindent In the previous subsection, we have considered the behaviour of the solitary wave solution under perturbations and have found 
the associated normal modes which guarantee stability. The perturbations with positive $\omega'$ decay asymptotically in space and oscillate harmonically in time, indicating the stability of the system. Instead of purely real energy, we can also have modes with complex $\omega'$ -- the so-called quasi-normal modes (QNMs) associated with purely outgoing boundary conditions. Such QNMs are damped oscillations in time
and have divergent behaviour in the spatial part of the wave function 
as $x'$ approaches the asymptotic infinities. Assuming the fact that collision of two solitary waves produces a perturbed solitary wave, which attains equilibrium by loosing energy, we can study the 
approach to stability of the resultant solitary wave through an
analysis of such QNMs.\\
A similar scenario is known in gravitational wave physics where QNMs play a major role in determining the parameters of the source and the remnant produced after a merger event \cite{vishveshwara, vishveshwara1,kokkotas, konoplya, cardoso}. QNMs of a BH are obtained where the cause of perturbation may not necessarily be known. Since QNMs do not depend on the cause of the perturbation, it is sufficient to study them  without bothering about the exact origin of
their occurence. Similarly, here we study the behaviour of a solitary
wave under perturbation and observe its approach to stability (in time) 
by finding the associated QNMs. The perturbed solitary wave may be 
produced as a result of a collision of two solitary waves or there could be other origins behind the fluctuations. One possible source of QNM excitation is the process of kink-antikink collision \cite{campos,Dorey}. Such collisions sometimes also harbour a resonance structure which have been studied for different models in \cite{campbell,resonance} . The conversion of a `shape (normal) mode' into a QNM has also been analysed. 
In our work we have primarily focused on the possible existence of quasi-normal modes, without providing a definite prescription or mechanism on how they may be excited.\\
Let us now proceed further on the analysis of the perturbations of our solitary wave solution, with the aim of knowing whether the confined harmonic oscillator potential supports QNMs. From known 
results on the standard square well potential \cite{squarewell, squarewell1, squarewell2, squarewell3, squarewell4}
for which {\em bound states, transmission resonances and quasinormal modes}
do exist we can expect that in our case we will also have all
three of them. Note that our perturbation potential differs
from the square well in the sense that the bottom of the potential
is not a constant in the domain $[-L',L']$, but is function of $x'$ (here, quadratic in $x'$). This is the reason why we expect results
similar to the square well.

\noindent We have used the transmission coefficient of the corresponding scattering problem to find the QNMs analytically and then verified them by numerically solving for QNMs using the direct integration approach. We must keep in mind that QNMs can be found by exactly solving the Schrodinger-like differential equation for only a handful of potentials \cite{triangular,visser}. Hence our confined harmonic potential is significant in its own right as the QNMs can be found exactly with the equations being solvable in terms of confluent hypergeometric functions, making the study even more worthwhile.

\subsubsection{\textbf{Time domain profiles}}

\begin{figure}[h]
 \centering
\begin{subfigure}[t]{0.47\textwidth}
  \centering
	\includegraphics[width=\textwidth]{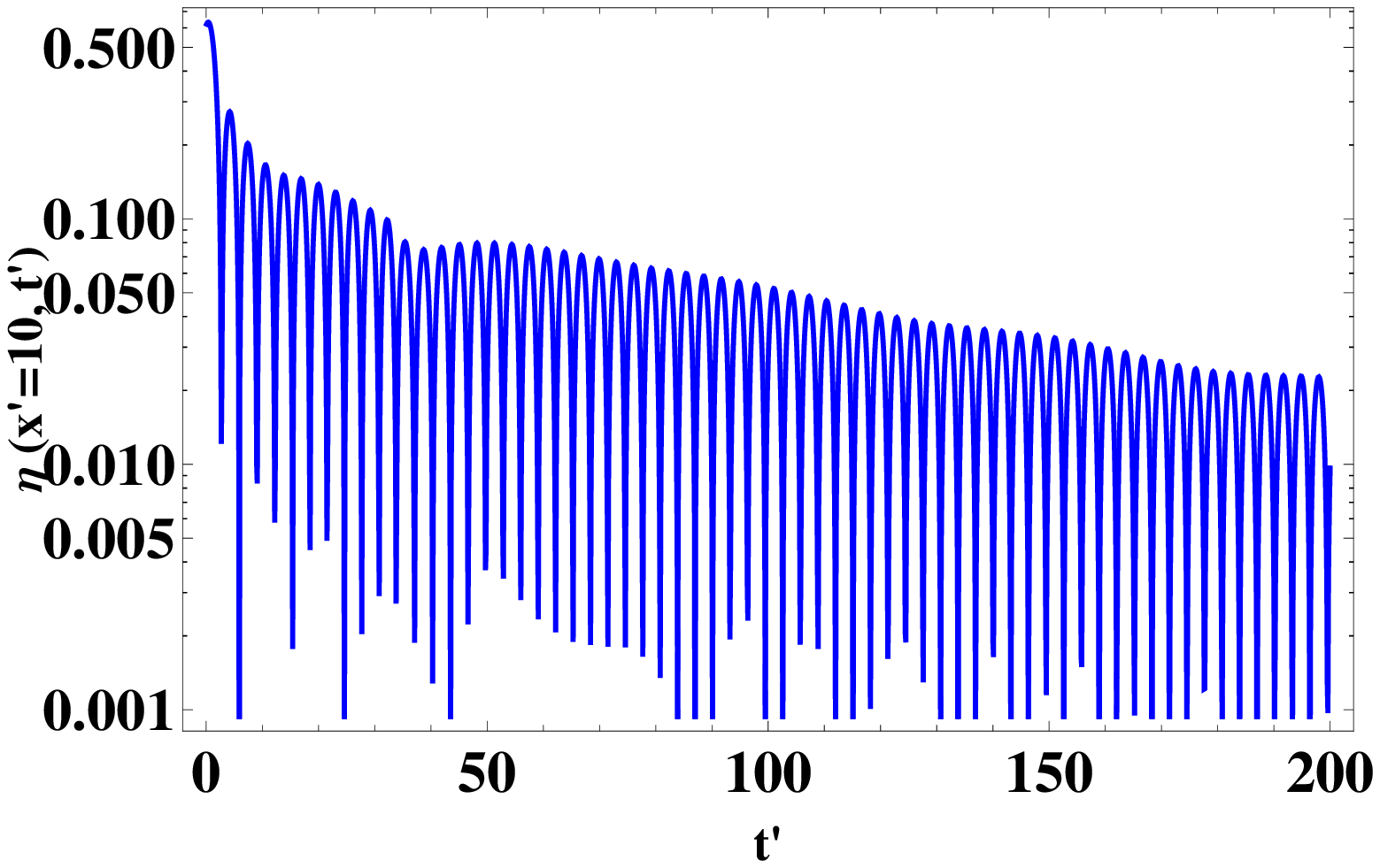}
 	\caption{ $ L'= 5, \alpha= 0.2\, (p=0.356825)$}
 	\label{fig:TD_1}
 \end{subfigure}
 \begin{subfigure}[t]{0.47\textwidth}
 	\centering
 	\includegraphics[width=\textwidth]{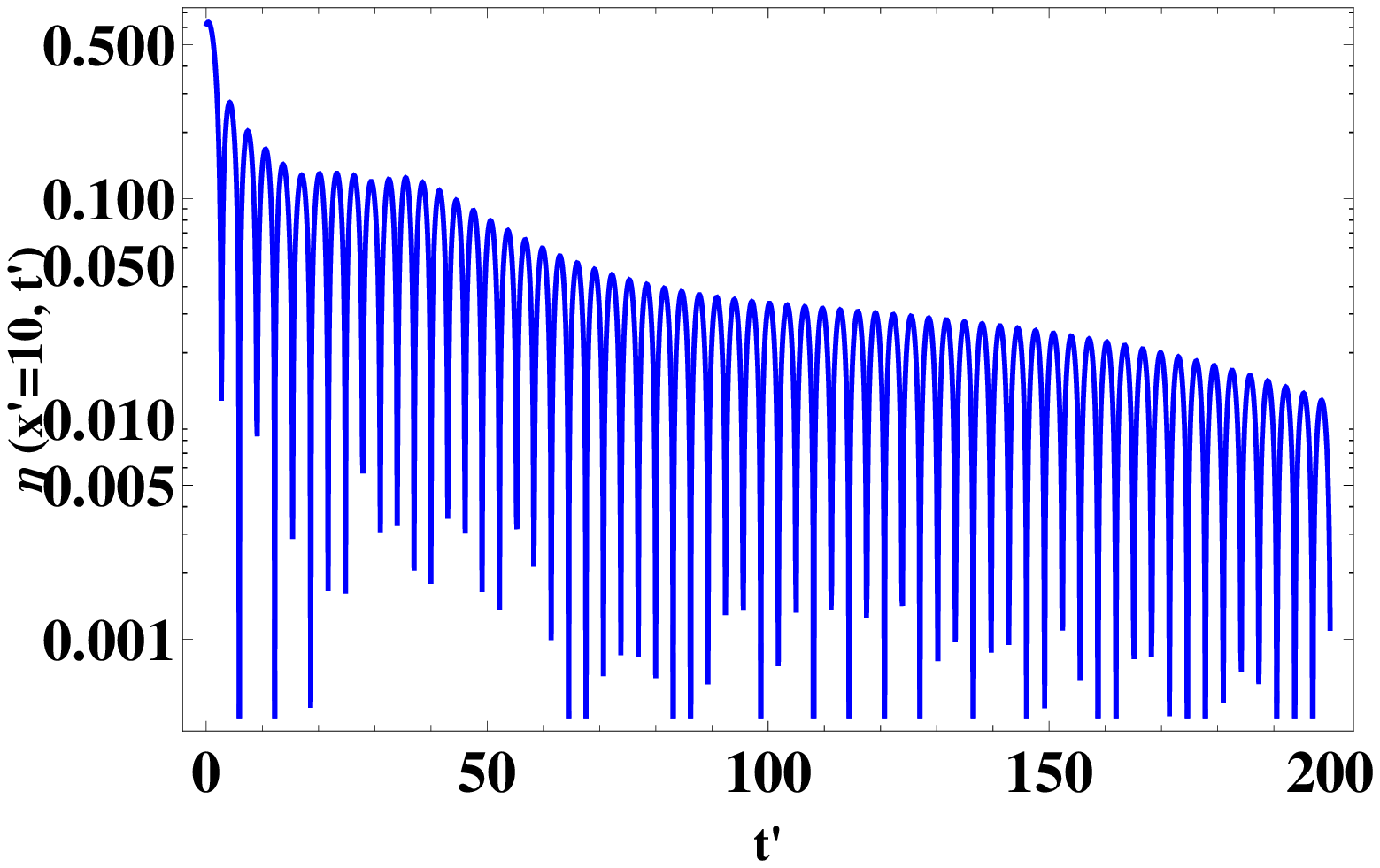}
 	\caption{$L'=3.97887, \alpha=0.251327 \,(p=0.4) $}
 	\label{fig:TD_2}
 	\end{subfigure}
 	\caption{Time domain profiles showing the characteristic QNM ringing in semi-logarithmic scale with  a Gaussian profile of the form $\eta(u,0) = e^{-(u-10)^2/100}$  as initial condition and evaluated at $x'=10$. Both the parameter value sets satisfy the solitary wave conditions.}
 	\label{fig:TD}
 \end{figure}

\noindent Following standard tools primarily used in 
graviatational wave physics, we recall that the time-domain profile 
provides a clear hint on the evolution of the perturbation as a function of time and the existence of QNMs. It also provides us with a way to check stability, a fact confirmed if the perturbation, at a particular spatial point, decays in time. In our case too, we 
demonstrate how $\phi (x',t')$ evolves as a function of time at a particular spatial location. Studying the perturbations of the solitary
wave $\phi(x',t')$ amounts to the study of the time-dependent perturbation $\eta(x',t')= \eta(x') e^{i \omega' t'}$.  We begin with the scalar wave equation given as
\begin{equation}
     \frac{\partial^{2}\eta(x',t')}{\partial t'^{2}} - \frac{\partial^{2}\eta(x',t')}{\partial x'^{2}} + V_{eff}(x') \eta(x',t') =0 
\end{equation}
where $V_{eff} =U(x')$ is the potential defined in eq.(\ref{U(x)}). This equation is converted to light-cone coordinates $du = dt' - dx'$ and $dv = dt' + dx'$, discretized and then integrated following the procedure as given in \cite{konoplya}  with a Gaussian profile of the form $\eta(u,0) = e^{-(u-10)^2/100}$ as the initial condition. We arrive at the time domain profiles shown in Fig.(\ref{fig:TD}) as observed at the spatial position $x'=10$ for two sets of parameter values satisfying the solitary wave condition (eq.(\ref{eq:continuity}) and eq.(\ref{eq:derivative})). 

\noindent The characteristic ringing pattern of the QNMs can be clearly observed in Fig.(\ref{fig:TD_1}) and (\ref{fig:TD_2}) for different parameter values of the potential. Thus we can conclude from the time domain profiles that our potential of eq.(\ref{U(x)}) does indeed harbour quasi-normal frequencies. The exact values of QNMs can be extracted using different techniques which we describe in detail in the forthcoming sections.\\
The field $\eta(x',t')$ decays slowly over a very long period of time. A small amount of back-scattering is also observed in the time evolution plots which leads to a slight increase in amplitude of the wave. This echo-like effect can be attributed to the non-zero value of the potential (which is unity in dimensionless variables) at spatial infinity which reflects back part of the wave. Even though our potential consists of a well, we do not observe any instability for any parameter values which is also evident from the time domain profiles. Thus, our results indicate the approach to stability of our perturbed solitary wave solution.

\noindent We move on now towards obtaining the quasi-normal modes.

\subsubsection{\textbf{QNMs from transmission coefficient}}

\begin{figure}[h]
      \centering
      \includegraphics[scale=0.55
      ]{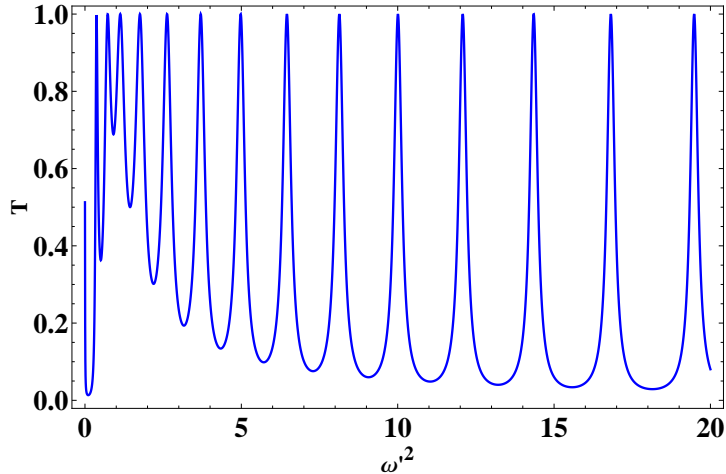}
      \centering
      \caption{Transmission resonance as observed for  $L'=5$, $\alpha = 0.2$}
      \label{fig:trans_resonance}
  \end{figure}

\noindent By definition, quasinormal modes obey purely outgoing boundary condition i.e. $\eta(x',t') \rightarrow e^{-i s |x'|}e^{i \omega' t'} $ with $s =  \sqrt{\omega'^2 -1}$ as $x'\rightarrow \pm \infty$. The quasinormal frequency, $\omega'$, is expected to have its imaginary component positive, so that the temporal part decays in time$(\sim e^{-Im(\omega') t'} )$ corresponding to a physically stable situation while the spatial part diverges($\sim e^{Im(s)|x'|}$) asymptotically. One way to computing such modes is to find the points where the complex transmission coefficient 
in the scattering problem for the Schr\"{o}dinger-like equation (\ref{Schrodinger}) diverges. For such a one-dimensional 
scattering problem  we have, 
\begin{equation}
\eta(x')  = \begin{cases} A e^{i s x'} +B e^{-i s x'} &; \,\,  -\infty<x' \leq -L' \\
 Cf(x')+Dg(x') &; \,\, -L' \leq x' \leq L' \\
 F e^{-i sx'} &; \,\,  L' \leq x'< \infty
\end{cases}
\end{equation}
where $f(x')$ and $g(x')$ are, as defined earlier (see eq.(\ref{hypergeometric})). Note that the definitions of scattering modes and transmission coefficient are a little different from the conventional way
(quantum mechanics), due to a different choice of time dependence($e^{i \omega' t'}$ instead of $e^{-i \omega' t'}$).  Here, $B e^{-i s x'}$ is the incoming wave from the left, approaching the potential. 
After some standard calculations, the transmission 
coefficient turns out to be,
\begin{equation}
\tilde{t}=\frac{F}{B}=i s e^{2 i s L'} \frac{f(L')g^{\prime}(L')-f^{\prime} (L') g(L')}{[i s g(L')+g^{\prime} (L')][i s f(L')+f^{\prime}(L')]}.
\end{equation}
The boundary conditions for QNMs demand only outgoing waves at spatial infinities, so the incident wave amplitude have to be set to zero in order
to obtain the QNMs. Thus, the transmission coefficient diverges 
since the denominator vanishes. The equation from which we obtain
the QNMs is therefore given as,
\begin{equation}\label{transcendental}
-(\omega'^2-1) f(L')g(L')+f^{\prime}(L')g^{\prime}(L')+ i  \sqrt{\omega'^2-1}[g(L')f^{\prime}(L')+f(L')g^{\prime}(L')]=0.
\end{equation}
Eqn.(\ref{transcendental}) is thus an analytical expression which contains all the
information on QNMs.
Only certain discrete values of $\omega'$ will satisfy the above transcendental equation which gives the exact QNM frequencies of the system.  To find the roots, we have plotted the real and imaginary parts of eq.(\ref{transcendental}) using \emph{ContourPlot} in \emph{Mathematica}. The intersection points are then used as initial values in \emph{FindRoot} which leads to the exact values of the QNM frequencies.\\
Apart from the QNMs, we can also observe the transmission resonances for real $\omega'^2$ in scattering (see Fig.(\ref{fig:trans_resonance})), where the transmission 
probability $T = |\tilde{t}|^2 $ reaches its maximum value i.e. one. For such frequencies, the potential well becomes totally transmitting with no reflection. Transmission resonances for various potentials are crucial in different contexts \cite{sksnm, chen, korsch}. However, 
here we will not discuss them any further. Instead, we
seek to verify the QNMs obtained using a different method, i.e. direct
integration.

\subsubsection{\textbf{QNMs from direct integration}}
\noindent In most physical scenarios, differential equations corresponding to the QNMs are difficult to solve analytically. Entirely numerical methods are the only way out. Numerical methods also provide a way of verifying the results obtained through analytical techniques as we will see in our case here. The direct integration method, also called the `shooting' method was 
first introduced by Chandrasekhar and Detweiler in \cite{Chandrasekhar}. Here, the Schr\"{o}dinger-like differential equation (eq.(\ref{Schrodinger})) is integrated straightaway, with proper boundary conditions, thereby yielding the QNMs. The method has been modified slightly 
while applying it to our specific potential.\\
The solution at infinity (or, in this case for $x'>L'$, i.e. positive $x'$) is exactly known in terms of exponential functions
\begin{equation}\label{asymptotic}
    \eta(x') = A_1\, e^{k x'} + A_2\, e^{-k x'}
\end{equation}
where $k = i \sqrt{\omega'^2-1}$. Ideally the integration should be done from $-\infty$ to $\infty$ but since our potential is symmetric about the origin, we integrate from $x'=0$ to $\infty$ and reflect the solution at the origin to arrive at the function in the region $x'<0$ (see \cite{aneesh,pdr}). There will be symmetric and anti-symmetric solutions corresponding to suitable matching conditions at the origin: namely, $\eta(0)=0$ for anti-symmetric and $\eta'(0)=0$ for symmetric solutions where $\eta'$ denotes derivative of $\eta$ with respect to $x'$. 

\noindent The differential equation (eq.(\ref{Schrodinger})) is then numerically integrated from the origin to $x'= L'$ with the initial conditions: $\eta(0)=0$ for the odd case and $\eta'(0)=0$ for even.
One has to keep in mind that the potential in the region $[-L',L']$ is that of a harmonic oscillator.\\
Next, we take the solution obtained after integration and equate it and its derivative with the solution obtained in eq.(\ref{asymptotic}) at $x'=L'$. This gives us expressions for $A_1$ and $A_2$ as functions of $\omega'$ which can be substituted back to eq.(\ref{asymptotic}) to get the general solution at any $x'>L'$. Finally, we make $A_1=0$ in the solution thus obtained, so that only outgoing waves exist at infinity (as per the convention used here, $A_1$ is associated with the incoming wave while $A_2$ involves the outgoing wave). The solution of this equation will give the values of the QNM frequencies. To check the consistency of our solution, we can calculate the ratio $A_1/A_2$ and see if it is nearly equal to zero. An added advantage of this method is that we can identify the QNMs as even or odd which was not possible in the transmission coefficient method discussed previously.
\subsubsection{\textbf{Comparing QNMs from the two methods}}
\noindent Finally we are in a position to implement the above-stated  methods and calculate the QNM frequencies for different parameter values of the potential. The following tables show QNM values for certain parameter choices as indicated there.
\begin{table}[h]
\begin{center}
\begin{tabular}{||c|c||}
     \toprule[0.8pt]
     $\omega'$ (From transmission coefficient) & $\omega'$ (From DI) \\ [1ex]
    \hline \hline
    0.966557 +i 0.0537962 & 0.966557 +i 0.0537962 \\ [1ex]
    1.59948 +i 0.321305 & 1.59948 +i 0.321305 \\ [1ex]
     2.22103 +i 0.420105 & 2.22103 +i 0.420105  \\ [1ex]
     \hline
    1.28732 +i 0.240357 & 1.28732 +i 0.240357 \\ [1ex]
    1.91022 +i 0.377235 & 1.91022 +i 0.377235  \\ [1ex]
    2.53224 +i 0.45485 & 2.53224 +i 0.45485  \\ [1ex]
   
    \bottomrule[0.8pt]
\end{tabular}
\caption{\label{tab:tab1}  Values of parameters are:    $\alpha= 0.2$ and $L'=5$. First three QNMs are odd while remaining three are even. }
\end{center}
\end{table}

\begin{table}[h]
\begin{center}
\begin{tabular}{||c|c||}
     \toprule[0.8pt]
     $\omega'$ (From transmission coefficient) & $\omega'$ (From DI) \\ [1ex]
    \hline \hline
    1.16485 +i 0.243292 & 1.16485 + i 0.243292 \\ [1ex]
   1.96892 +i 0.464461 & 1.96892 +i 0.464461  \\ [1ex]
    2.75861 +i 0.573992 & 2.75861 +i 0.573992  \\ [1ex]
     \hline
    1.57139 +i 0.380764 & 1.57139 +i 0.380764 \\ [1ex]
    2.3641 +i 0.52566 & 2.3641 +i 0.52566  \\ [1ex]
    3.15296 +i 0.613966 & 3.15296 +i 0.613966    \\ [1ex]
   
    \bottomrule[0.8pt]
\end{tabular}
\caption{\label{tab:tab2}  Values of parameters are:    $\alpha= 0.251327$ and $L'=3.97887$. First three QNMs are odd while remaining are even. }
\end{center}
\end{table}

\begin{figure}[h]
 \centering
\begin{subfigure}[t]{0.47\textwidth}
  \centering
	\includegraphics[width=\textwidth]{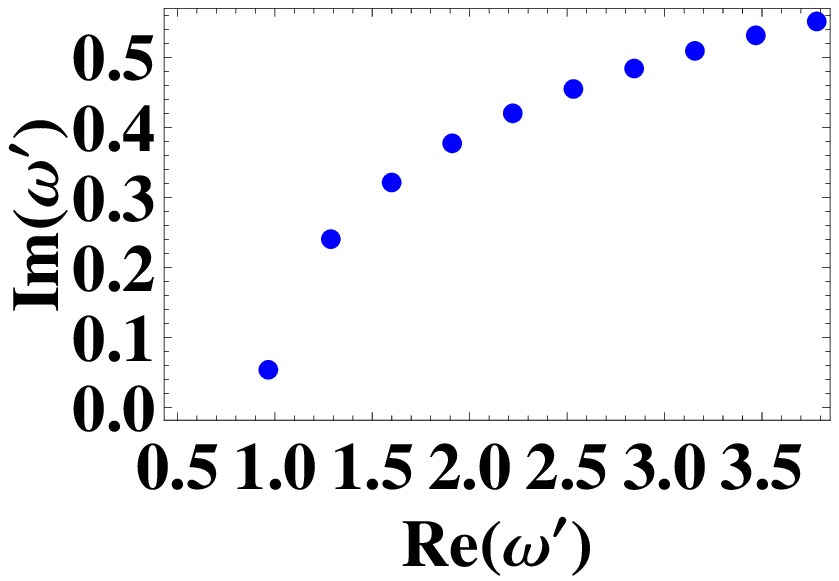}
 	\caption{ $ L'= 5, \alpha= 0.2$}
 	\label{fig:para_1}
 \end{subfigure}
 \begin{subfigure}[t]{0.47\textwidth}
 	\centering
 	\includegraphics[width=\textwidth]{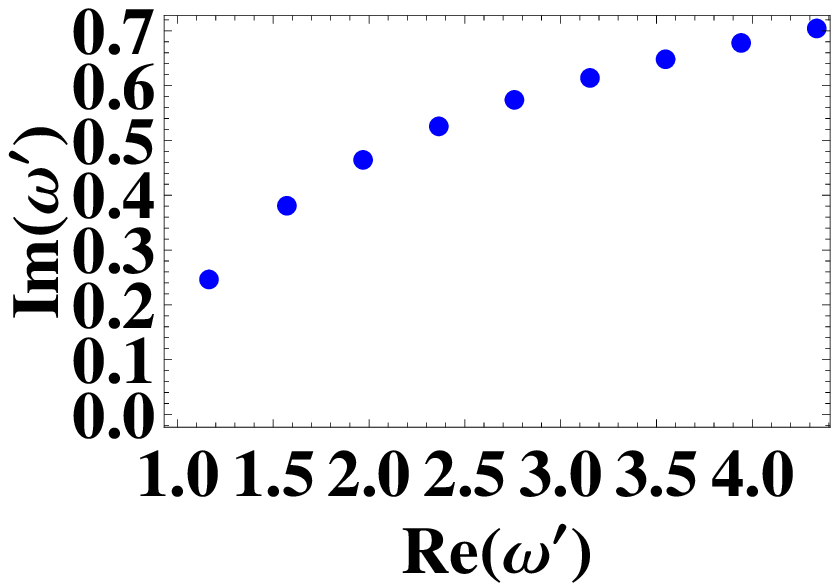}
 	\caption{$ L'=3.97887, \alpha=0.251327 $}
 	\label{fig:para_2}
 	\end{subfigure}
 	\caption{Variation of real and imaginary part of QNM frequencies for two sets of parameter values that satisfy the soliton criteria.}
 	\label{fig:soliton_para}
 \end{figure}
\noindent From the tables \ref{tab:tab1} and \ref{tab:tab2}, it is quite evident that the results from both the methods are consistent with each other. The parameter values chosen to obtain the results in 
Table \ref{tab:tab1} and Table \ref{tab:tab2} satisfy solitary wave conditions mentioned  earlier in eq.(\ref{eq:continuity}) and (\ref{eq:derivative}). In Fig.(\ref{fig:soliton_para}), the variation of real and imaginary parts of the QNM frequencies can be seen. All QNM frequencies have a positive imaginary part indicating the decay of perturbation with time. Hence we can conclude that the potential in eq.(\ref{U(x)}) is stable against perturbations both for an independent set of parameters as well as for our solitary wave solution. The results from DI match exactly, up to the decimal places considered here, with the solutions obtained using the transmission
coefficient, because in DI we have not used any series expansion and all exact solutions have been incorporated. Obviously if we compare the QNM values from the two methods after the decimal places shown in the tables, we do find differences. One might also notice that even though the time domain profiles indicate a slow decay of signal, we find the imaginary parts of the QNM frequencies to be not so small as to produce such a large decay time. One possible explanation for the slow decay might be attributed to the small reflections from a non-zero $V_0$ or in the dimensionless variables notation, unity, at spatial infinity which induces a greater strength in the signal. Due to this additional reflected signal we do not observe a sharp damping as is usually expected in QNM frequencies. Thus it can be concluded that the time domain profile is not dominated by any individual QNM frequency of the potential.
We can further verify this fact if we apply the Prony method \cite{konoplya} and extract the dominant frequency directly from the time evolution. This can be considered as an example of the observation made by Nollert \cite{nollert} for discontinuous potentials, where it is said that even though the QNMs form a complete set, there is no single mode among the QNM frequencies that dominates  time evolution. \\
Apart from the study of the confined harmonic oscillator potential in the context of solitary wave solution we can investigate features associated with the given $U(x')$, as an independent entity. A brief discussion on how the QNM frequencies depend on the parameters in $U(x')$ can be found in the Appendix. Obviously without the 
solitary wave criteria, we can freely vary the parameters $\alpha,$ $L'$ and see their effects. Such a study could be useful in other areas where a potential like $U(x')$ may arise. 

\section{\bf Conclusion}

\noindent Solitary waves have interested researchers from
various branches of physics over many years. The property of 
such solutions retaining their shape while traveling,
renders them as coherent structures created due  to the 
presence of nonlinearity in the equations of motion. 
Finding a model non-linear theory with dispersion and no dissipation 
with solitary wave solutions is not an easy task. 
In our work we have developed a new model with a solitary wave 
solution in a massless scalar field theory in 1+1 dimensions. 
The functional dependence of the potential on the scalar field $\phi$ was not defined to begin with. We constructed the potential $V(\phi)$ such that (a) our theory supports a non-trivial solitary wave solution and  
(b) the equation for 
fluctuations about the solitary wave is analytically solvable
(in our case it is just a confined simple
harmonic oscillator). We also checked the localisation of energy density and the finiteness of total energy for our solitary waves. 
In our presentation, we have used dimensionless variables $(x',t')$ and have 
shown how to switch over to
dimensionful ones, when needed.

\noindent After obtaining the soliton solution we moved on to studying its stability by finding the normal and quasi-normal modes of scalar  perturbations. As stated before, the potential $V(\phi)$ was constructed such that the perturbation potential $U(x')$ is in the form of a confined harmonic oscillator. That the potential is stable against perturbation was evident from the normal modes. With the calculation of QNMs we also verified the stability in time through decaying perturbations. We have obtained the QNMs analytically through the divergence of the transmission coefficient and also numerically, through direct integration. Both methods provide QNM values that match perfectly upto the decimal places considered. Finally, an interesting observation from the QNM study is the fact that no single QNM frequency dominates the time domain profile which can be attributed to the discontinuity of the potential. 

\noindent In summary, our work presents a novel scalar field model with
solitary waves for which the fluctuation equation is 
easy to analyse quantitatively. For most known solitary waves and solitons, one ends up with a fluctuation equation with a rather complicated potential, which is tedious to solve. Here, we have
a model for which the solitary wave arises through a tailor-made
potential with the perturbation equation simple and easy to
solve and analyse. We believe there could be many similar models
wherein both the solitary waves as well as its perturbations 
are analytically tractable, which makes the 
quantitative understanding of the system relatively easier. 
Scenarios where double-well potentials generically arise are viable candidates 
where our model can be of use, possibly in a physically relevant
context. Finally, newer models explaining how  quasinormal modes
may be excited in the context of solitary waves, need to be built, in order to provide the right context
wherein the possibilities (on QNMs in particular) discussed here, in this work, will be
of importance, in future. 

\section*{Acknowledgements}

\noindent Surajit Basak  thanks Centre for Theoretical Studies, IIT
Kharagpur, India for informal visits during 2019 when this work was
initiated and carried out. He also thanks his present host, P. Piekarz, Institute of Nuclear Physics, Polish Academy of Sciences, Krakow, Poland for allowing him to use his present address, 
in this article.

\appendix

\section{\textbf{Variation of QNMs with parameters}}

\begin{figure}[h]
\begin{subfigure}[t]{0.49\textwidth}
  \includegraphics[width=\textwidth]{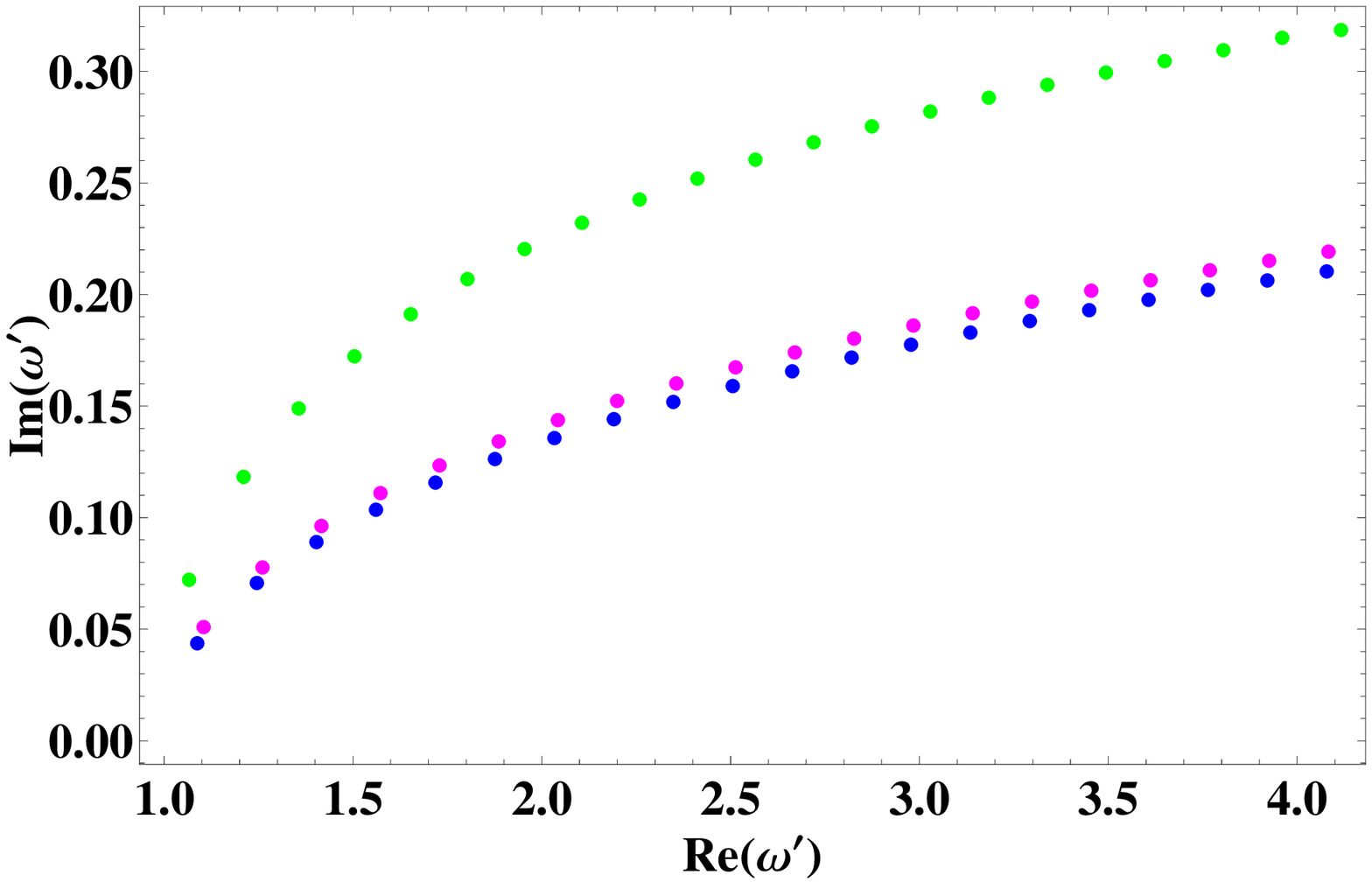} 
    \caption{$L'=10$, $\alpha = 0.025$ (\tikzcircle[blue, fill=blue]{1.5pt}), 0.05 (\tikzcircle[magenta, fill=magenta]{1.5pt}), 0.1(\tikzcircle[green, fill=green]{1.5pt}) }
    \label{fig:alpha_vary}
\end{subfigure}
\begin{subfigure}[t]{0.49\textwidth}
  \includegraphics[width=\textwidth]{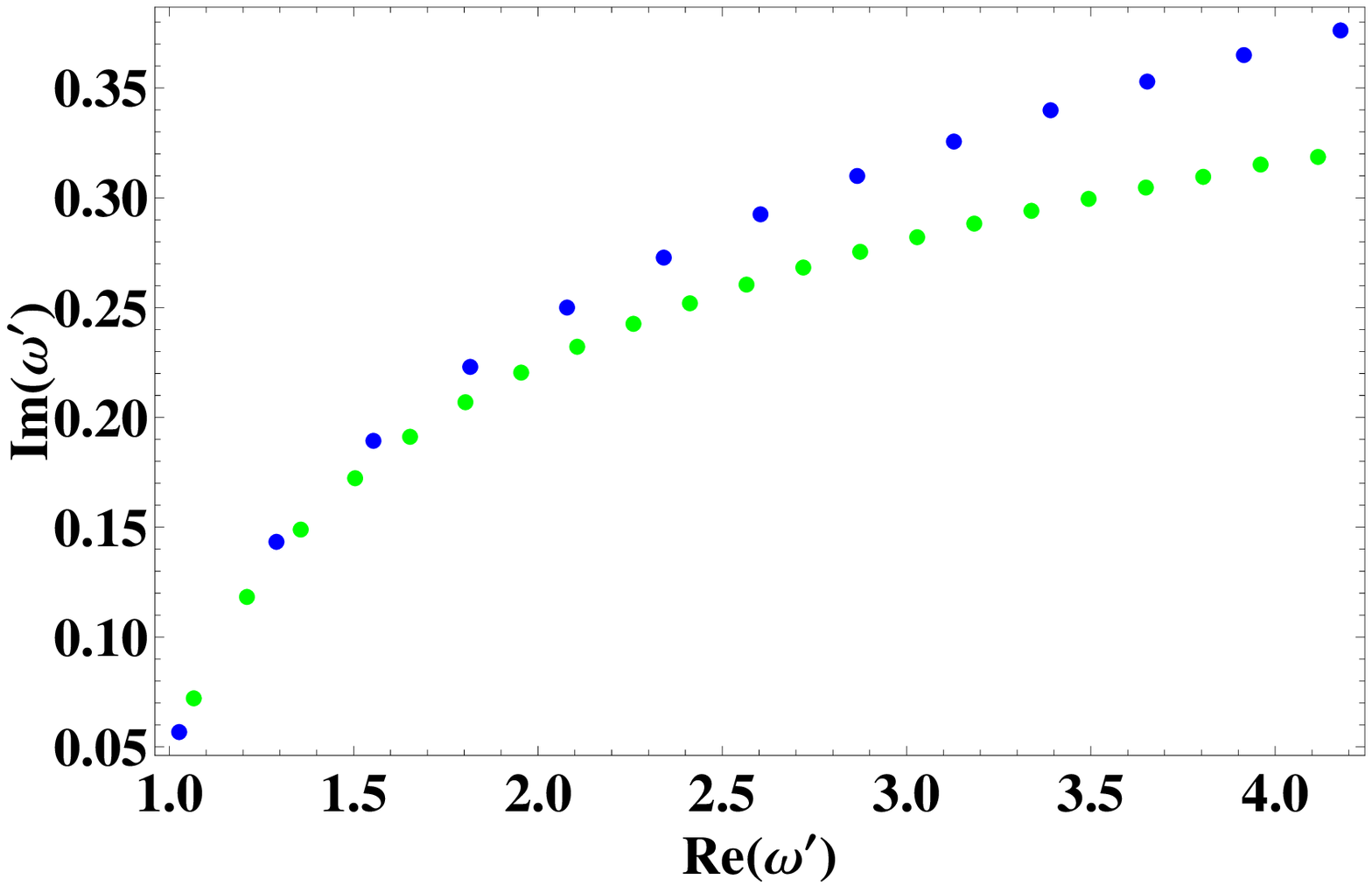} 
   \caption{$\alpha=0.1,L'=$6(\tikzcircle[blue, fill=blue]{1.5pt}),10(\tikzcircle[green, fill=green]{1.5pt})}
    \label{fig:L_vary}
\end{subfigure}

\caption{ $Im(\omega')$ vs $Re(\omega')$ plotted for different parameter values.}
\label{fig: parameter_vary}	
\end{figure}

\noindent In this appendix we discuss the dependence of the QNM frequencies on the parameters of the potential. Such an analysis will help us in appreciating the confined harmonic potential $U(x')$ as an independent, discontinuous potential with its parameters not necessarily obeying the solitary wave criteria. This independent study could be of use in scenarios
where a similar potential may arise.
We note that the potential $U(x')$ involves parameters: $\alpha$ and $ L'$, which are independent if we do not demand its solitary wave connect. Hence, we may vary them freely and see how the QNM frequencies get affected.
In Fig.(\ref{fig: parameter_vary}) we show how the spectrum of the QNMs obtained 
differ when we vary each parameter separately.

\noindent $\bullet$ In Fig.(\ref{fig:alpha_vary}) we observe the effect of variation of $\alpha$ on the QNMs. The value of $\alpha$ determines the depth of the potential well.  One must keep in mind while choosing the parameters, that we are allowed to take only those values obeying $\alpha(\alpha L'^2-1) < 1$ (
this guarantees that the nature of the confined harmonic oscillator potential is preserved). As $\alpha$ increases, the well becomes narrow and the imaginary part of the QNM frequency increases.\\
\noindent $\bullet$ Changing the magnitude of discontinuity or jump at $x'= \pm L'$ also affects the QNM spectrum as observed in Fig.(\ref{fig:alpha_vary}). If we go to lower values of $\alpha$, the well becomes shallower while its width ($2L'$) remains fixed, which results in a larger discontinuity. Hence we see that small values of $\alpha$ correspond to lower values of $\omega'_i$. \\
\noindent $\bullet$ When $L'$ is varied i.e. the width of the potential  increases (see Fig.(\ref{fig:L_vary})), the imaginary part of QNM decreases. For very large $L'$, the imaginary part will keep on getting smaller. Finally for $L'\rightarrow \infty$ the well vanishes leaving a constant potential with no interesting features. \\
\noindent In general, we find the lower modes to remain almost unaffected by a change of the parameters. Also, the real part of QNMs remain nearly same under parameter variation, whereas, it is the imaginary part which shows the effect. Working in dimensionful variables would lead to the introduction of $V_0$ (in inverse length squared units) in the potential. Changing the magnitude of $V_0$ with the other parameters kept fixed will affect the discontinuity. Hence increasing $V_0$ or decreasing $\alpha$ will have the same effect on the QNM spectrum.\\

\end{document}